# On a pragmatic approach optical analogues of gravitational attractors.

D. P. San-Roman-Alerigi [1], A. B. Slimane [1], T. K. Ng [1], M. Alsunaidi [1,2], and B. S. Ooi [1*]

[1]Photonics Laboratory, Physical Sciences and Engineering Division, King Abdullah University of Science and Technology (KAUST), Saudi Arabia.
[2]King Fahd University of Petroleum and Minerals (KFUPM), Saudi Arabia.

*corresponding author: boon.ooi@kaust.edu.sa

## Abstract

*In our work we theoretically demonstrate a refractive index mapping to enable optical analogues to celestial mechanics, where is possible to achieve light confinement and trapping by means of a static, and planar, refractive index mapping which could be implemented under current technological and [meta]material constraints at optical frequencies. The mathematical and physical background to make possible these effects bring forth an exciting ground to test celestial mechanics in the laboratory, and provides the key to enable miscellany of planar optical system that are of great interest to photonic applications, namely optical time delays, transient optical memories and random resonators.*

Light trapping and manipulation is an enthralling and fundamental research, which can enable novel photonic applications like light-time delays and optical memories; while providing an exciting framework to mimic celestial mechanics of interest to fundamental physics.

Applications and experimental frameworks alike, arise from the relation between geometry and refractive index of materials, which has been studied for several decades through diverse methods, most based on differential geometry [1-4]. They, have been key to the theoretical realization of optical cloaks, optical illusion devices and recently black-hole analogues in moving refractive index perturbations [3-11]. Concerning light trapping different methods have been proposed, recently Genov *et al* presented an isotropic inhomogeneous media which was able to confine light mimicking a closed orbit around a black hole [12]. However, the electromagnetic properties, permittivity and permeability, presented a singularity which could be difficult to realize in known [meta]materials; furthermore implement the mapping under current foundry technologies could be challenging.

To avoid singularities, and limit the refractive index to feasible values, in our study we solve the inverse full-relativistic eikonal system of equations (equations (1) - (4)), subject to materials constraints [13]. The aforementioned equations result from solving the Lagrange-Euler problem, where the potential is given by the refractive index, and the 4-dimensional dynamic properties of light propagation are linked by the affine parameter $\tau$, *i.e.* the proper time. The inverse problem is initiated by setting the trap orbits *a priori*. The result is a radial, static, planar, and real-valued refractive index map.

$$r'' = r\phi'^2 + \frac{t'^2\, \partial_r n}{n^3}, \qquad (1)$$

$$\phi'' = \frac{-2rr'\phi' + \frac{t'^2\, \partial_\phi n}{n^3}}{r^2}, \qquad (2)$$

$$z'' = \frac{t'^2\, \partial_z n}{n^3}, \qquad (3)$$

$$t'' = \frac{t'\left(t'\, \partial_t n + 2(z'\, \partial_z n + \phi'\, \partial_\phi n + r'\, \partial_r n)\right)}{n}, \qquad (4)$$

The refractive index map corresponding to the trapping orbits are shown in figures 1 and 2. The refractive index follow a Gaussian-like distribution, where $n_A$ and $n_C$ represents the maximum and minimum refractive index values, respectively. As it can be seen in figure 2, the radial and angular velocity components of the trapped photon are periodic functions, whose rate depends on the refractive index characteristics. The maximum and minimum of the refractive index map depend on the minimum and maximum trapping radius, as is observed in figure 1.

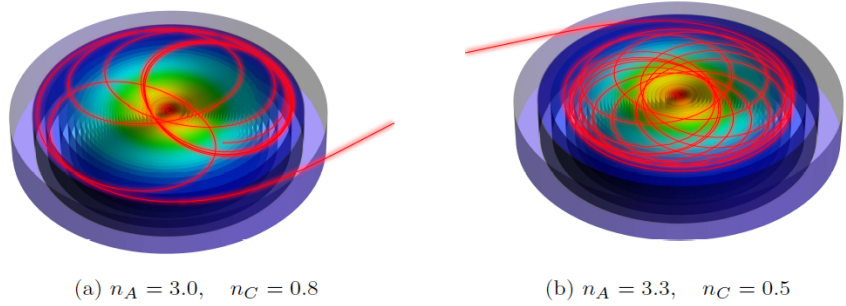

(a) $n_A = 3.0$, $n_C = 0.8$      (b) $n_A = 3.3$, $n_C = 0.5$

**Figure 1**: *Light trapping dynamics for Gaussian refractive index shape for different $n_A$ and $n_C$. The maximum value of $n$ in all cases is 3.8 at the center of the device. $\sigma$ is constant for all results. Notice the increment in the internal radius $r_{min}$ as $n_C$ increases, and the trapping orbit width $w_o$ reduces.*

We found that trapping is highly sensitive to the refractive index distribution perturbations, a change of $\Delta n > 0.05$ modifies entirely the trapping orbit. Yet small perturbations, $\Delta n < 0.01$, do not change the overall behaviour of the trap. This sensitivity is an interesting opportunity. and a key feature, to explore dynamic trapping, both important to optical memory and time delays.

A prototype Gaussian trap could be fabricated in by stacking different photo-refractive [meta]materials in multiple layers. Recall that in photo-refraction the optical properties of the material, permeability and permittivity, change as a function of the light intensity, and hence it is possible to modify the refractive index in time, thence enabling or disabling the trap at will.

Is worth noting, that the mathematical and physical treatment of the problem bring forth an exciting ground to test celestial mechanics in the laboratory; and provides the key to enable miscellany of planar optical system that are of great interest for diverse photonic applications, *e.g.* time delays, temporal optical memories, or random resonators, to name a few; thence its importance.

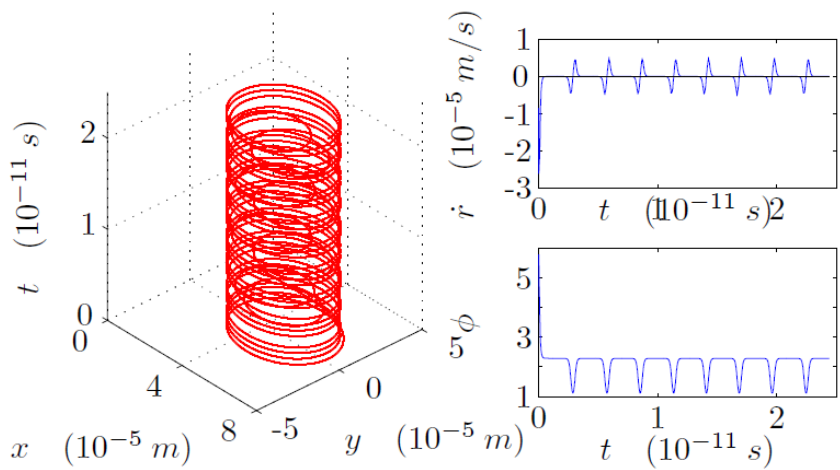

**Figure 2**: *Light trapping solutions for a radial varying refractive index $n(\vec{x}) = n(r) = exp^{-r^2/\sigma^2}$.*

# References

[1] U. Leonhardt and T. G. Philbin, "Transformation optics and the geometry of light," Progress in Optics, vol. 53, pp. 69–152, 2008.

[2] H. Chen, C. T. Chan, and P. Sheng, "Transformation optics and metamaterials.," Nature materials, vol. 9, pp. 387–96, May 2010.

[3] J. B. Pendry, D. Schurig, and D. R. Smith, "Controlling electromagnetic fields.," Science, vol. 312, pp. 1780–2, June 2006.

[4] D. Schurig, J. B. Pendry, and D. R. Smith, "Calculation of material properties and ray tracing in transformation media.," Optics express, vol. 14, pp. 9794–804, Oct. 2006.

[5] U. Leonhardt and P. Piwnicki, "Optics of nonuniformly moving media," Physical Review A, vol. 60, pp. 4301–4312, Dec. 1999.

[6] U. Leonhardt and P. Piwnicki, "Relativistic Effects of Light in Moving Media with Extremely Low Group Velocity," Physical Review Letters, vol. 84, pp. 822–825, Jan. 2000.

[7] I. Brevik and G. Halnes, "Light rays at optical black holes in moving media," Physical Review D, vol. 65, Nov. 2001.

[8] R. Liu, C. Ji, J. J. Mock, J. Y. Chin, T. J. Cui, and D. R. Smith, "Broadband ground-plane cloak.," Science, vol. 323, pp. 366–9, Jan. 2009.

[9] T. Ergin, N. Stenger, P. Brenner, J. B. Pendry, and M. Wegener, "Three-dimensional invisibility cloak at optical wavelengths.," Science (New York, N.Y.), vol. 328, pp. 337–9, Apr. 2010.

[10] Y. Lai, J. Ng, H. Chen, D. Han, J. Xiao, Z.-Q. Zhang, and C. Chan, "Illusion Optics: The Optical Transformation of an Object into Another Object," Physical Review Letters, vol. 102, pp. 1–4, June 2009.

[11] D.-H. Kwon and D. H. Werner, "Flat focusing lens designs having minimized reection based on coordinate transformation techniques.," Optics express, vol. 17, pp. 7807–17, May 2009.

[12] D. Genov, S. Zhang, and X. Zhang, "Mimicking celestial mechanics in metamaterials," Nature Physics, vol. 5, pp. 687–692, July 2009.

[13] D. P. San-Roman-Alerigi, T. K. Ng, M. Alsunaidi, Y. Zhang, and B. S. Ooi, "Generation of j0-bessel- gauss beam by an heterogeneous refractive index map," Journal of the Optical Society of America A, vol. accepted, March 2012